\documentclass[%
  reprint,
  superscriptaddress,
  amsmath,amssymb,
  aps,
  floatfix,
]{revtex4-2}

\usepackage{graphicx}
\usepackage{dcolumn}
\usepackage{bm}
\usepackage[colorlinks=true, allcolors=blue]{hyperref}

\usepackage{dsfont}
\usepackage{xcolor}
\usepackage{todonotes}
\usepackage{listings}
\usepackage{physics}
\usepackage{float}
\usepackage{siunitx}

\usepackage{caption}[justification=justified]
\usepackage{subcaption}[justification=justified]

\definecolor{fedecolor}{HTML}{FEDEBE}
\definecolor{fedecolor2}{HTML}{ff7600}

\definecolor{pinocolor}{HTML}{991111}

\begin{document}

\preprint{APS/123-QED}

\title{Synchronization of quantum communication over \\an optical classical communication channel}

\author{Federico~Berra}
\email{federico.berra@phd.unipd.it}
\affiliation{%
 Dipartimento di Ingegneria dell'Informazione, Universit\`a degli Studi di Padova, via Gradenigo 6B, 35131 Padova, Italy
}%

\author{Costantino~Agnesi}
\affiliation{%
 Dipartimento di Ingegneria dell'Informazione, Universit\`a degli Studi di Padova, via Gradenigo 6B, 35131 Padova, Italy
}%

\author{Andrea~Stanco}
\affiliation{%
 Dipartimento di Ingegneria dell'Informazione, Universit\`a degli Studi di Padova, via Gradenigo 6B, 35131 Padova, Italy
}%

\author{Marco~Avesani}
\affiliation{%
 Dipartimento di Ingegneria dell'Informazione, Universit\`a degli Studi di Padova, via Gradenigo 6B, 35131 Padova, Italy
}%

\author{Michal~Kuklewski}
\affiliation{%
 Thales Alenia Space Schweiz AG, Schaffhauserstrasse 580, CH-8052 Z\"urich, Switzerland
}%

\author{Daniel~Matter}
\affiliation{%
 Thales Alenia Space Schweiz AG, Schaffhauserstrasse 580, CH-8052 Z\"urich, Switzerland
}%

\author{Paolo~Villoresi}
\affiliation{%
 Dipartimento di Ingegneria dell'Informazione, Universit\`a degli Studi di Padova, via Gradenigo 6B, 35131 Padova, Italy
}%
\affiliation{%
 Padua Quantum Technologies Research Center, Universit\`a degli Studi di Padova, via Gradenigo 6B, 35131 Padova, Italy
}%

\author{Giuseppe~Vallone}
\affiliation{%
 Dipartimento di Ingegneria dell'Informazione, Universit\`a degli Studi di Padova, via Gradenigo 6B, 35131 Padova, Italy
}%
\affiliation{%
 Padua Quantum Technologies Research Center, Universit\`a degli Studi di Padova, via Gradenigo 6B, 35131 Padova, Italy
}%
\affiliation{%
 Dipartimento di Fisica e Astronomia, Università degli Studi di Padova, via Marzolo 8, 35131 Padova, Italy
}%


\date{\today}

\begin{abstract}
Precise synchronization between transmitter and receiver is crucial for quantum communication protocols, such as Quantum Key Distribution (QKD), to efficiently correlate the transmitted and received signals and increase the signal-to-noise ratio.
In this work, we introduce a synchronization technique that exploits a co-propagating classical optical communication link and test its performance in a free-space QKD system. 
Previously, existing techniques required additional laser beams or relied on the capability of retrieving the synchronization from the quantum signal itself, though this is not applicable in high channel loss scenarios. On the contrary, our method exploits classical and quantum signals locked to the same master clock, allowing the receiver to synchronize both the classical and quantum communication links by performing a clock-data-recovery routine on the classical signal. In this way, by exploiting the same classical communication already required for post-processing and key generation, no additional hardware is required, and the synchronization can be reconstructed from a high-power signal. Our approach is suitable for both satellite and fiber infrastructures, where a classical and quantum channel can be transmitted through the same link.
\end{abstract}

\maketitle

\section{Introduction}
\begin{figure*}
  \includegraphics[width=\linewidth]{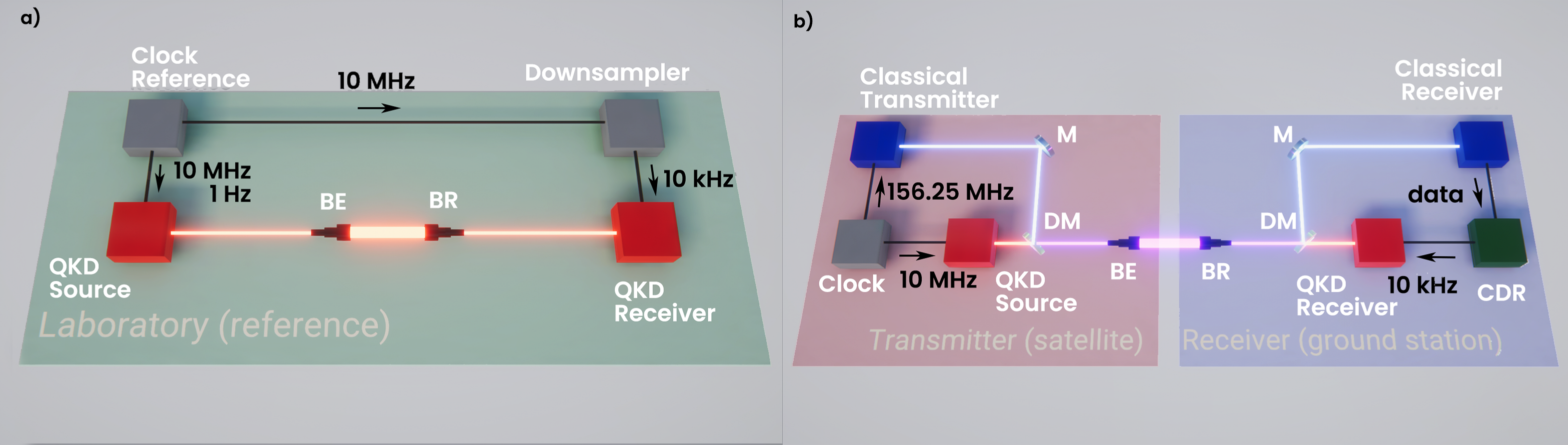}
  \caption{\textbf{a)} In this figure, a QKD setup designed for testing in the laboratory is represented. The QKD source and receiver (red) are synchronized by a clock reference (gray).
  \textbf{b)} In this figure, a QKD setup designed for space is represented. The transmitter (light-red) is composed of: clock sources (gray), classical transmitter (blue), QKD source (red), metallic mirror (M), dichroic mirror (DM), and beam expander (BE). The receiver (light-blue) is composed of: a beam reducer (BE), dichroic mirror (DM), metallic mirror (M), QKD receiver (red), classical receiver (blue), and clock data recovery (green).
}
  \label{fig:SetupsSchemes}
\end{figure*}
Modern satellite systems and constellations are increasingly adopting optical communications over traditional Radio Frequency (RF) links. This is mainly motivated by three key advantages of optical links over RF systems. First, optical communications offer a higher bit rate and a workable communication distance than traditional RF with lower volume, mass, and power requirements. Second, the low divergence and strong directionality of optical links offer increased channel security. Lastly, the lack of frequency license regulations is particularly advantageous since the allocation of RF bands is close to saturation. This increased interest in optical communications has promoted the development of test systems and demonstrations have been carried out for a wide variety of scenarios including: Low Earth Orbit (LEO)-to-LEO~\cite{Fields2009}, LEO-to-ground~\cite{Gregory2013,Casado2017}, Geostationary Equatorial Orbit (GEO)-to-LEO~\cite{Zech2015}, GEO-to-ground~\cite{Alonso2004}, and even deep space-to-ground~\cite{Borson2014}. Additionally, several projects are under development for future demonstrations~\cite{Sansone2020}.

Similarly, Quantum Communications (QC) is seeing increased interest and deployment in terrestrial and satellite communication networks. This is mainly due to QC's ability to solve many open problems in classical communications by exploiting quantum resources such as entanglement and single photons. In particular, Quantum Key Distribution is the most mature QC protocol~\cite{Pirandola2020} and has strategic importance since it allows two parties to generate a cryptographic key whose security is guaranteed by the principles of quantum mechanics~\cite{Scarani2008}. In this regard, Satellite-based QKD systems are of crucial importance in the development of global-scale QC networks since they allow to overcome distance limitations imposed by the exponential losses in optical fibers~\cite{Boaron2018}. In August 2016, the Chinese Academy of Science launched Micius, a dedicated satellite for quantum communications tasks including QKD~\cite{Liao2017_Sat}, achieving key exchanges at an intercontinental scale~\cite{Liao2018}. Regardless of these pioneering results, satellite-based QKD is still far from widespread adoption and several technological efforts must be carried out to achieve technical and economical feasibility.

A promising approach that could result in lower costs of QKD systems and improve their adoption in satellite systems would be enhancing optical communication links with QKD. In this way, the QKD device could share many essential subsystems with the optical communication unit, such as the optical head and the Pointing, Acquisition, and Tracking (PAT) systems. Furthermore, the communication channel that is implemented via optical communication could be exploited to assist the QKD in the post-processing steps that are crucial for secure key generation.

In this work, we extend the interconnection between the satellite QKD system and the optical communication system by studying the possibility of synchronizing the QKD transmitter and receiver using the clock-data-recovery (CDR) of the optical link exclusively. Since, traditionally, most QKD systems require an additional link to share the QKD clock, our approach reduces the overall complexity and hardware required for satellite-based QKD. 

\section{Setup}
Synchronization between transmitter and receiver is crucial in QKD and, in general, in any quantum communication experiment such as the distribution of entanglement.
QKD is implemented by sending a train of qubits, namely short ($\sim1 ns$) optical pulses attenuated to the single-photon level which are then encoded in a suitable degree of freedom. Synchronization then allows the receiver to know the expected time of arrival of the qubits and to discard the detections that are generated by the background noise or the detector's dark counts by defining a suitable detection window. This filtering procedure is crucial for the success of Quantum Communication protocols, especially when compared to classical communication, mainly due to the single-photon nature of the signal and the discrete nature of single-photon detectors.  
Achieving narrow detection windows increases the signal-to-noise ratio and performance. Typical detection windows for QKD are of the order of a few nanoseconds~\cite{Agnesi:19}, and their effective widths depend on the ability of synchronization and the intrinsic jitter of the receiver apparatus.

It is worth noting that the start time of the communication and the repetition rate are not enough to fix the reference system due to the drifts of physical clocks. To compensate for drifts in classical communication, it is possible to recover the transmitter's frequency directly from the received signal with a clock-data-recovery algorithm (CDR).
Qubit-based synchronization algorithms have also been developed for quantum communication~\cite{Calderaro2020}, but the performances of such procedures are non-optimal in the presence of fading quantum channels. 

An alternative solution for QKD is to phase lock the source and the receiver with an external clock as shown in Fig.\ref{fig:SetupsSchemes}-a. The clock signal is then shared between the two parties directly or by using a dedicated optical link that exploits a pulsed laser and a photodiode.  This solution is already used in many setups~\cite{Korzh2015,Dynes2016,Liu19} and may require clocks with different repetition rates (locked in phase) to match the instrumental requirements between source and receiver.

Here, we present a synchronization technique that instead exploits a laser communication link that co-propagates with the QKD signal. 
The setup is shown in Fig.\ref{fig:SetupsSchemes}-a and is designed to satisfy the prepare-and-measure quantum key distribution protocol. The transmitter is composed of three subsystems: a clock reference, a classical transmitter, and a QKD source. The clock reference locks in phase the QKD source with a standard telecommunications free-space transmitter. Its optical output in the C-band is merged by a dichroic mirror with the output of the QKD source. 

The discrete-variable QKD source is realized with a laser source and a polarization modulator controlled by an FPGA-based system implemented on a ZedBoard (Zynq7020) evaluation board~\cite{Stanco2022}. The laser at $800 nm$ generates a train of pulses with repetition rates of $50 MHz$. The pulses are then converted into quantum symbols by intensity~\cite{Roberts2018} and polarization modulation~\cite{Avesani2020}, implementing a three-states one-decoy BB84 protocol~\cite{Grunenfelder2018, Agnesi2020}.
The optical signal is then transmitted through a telescope to the receiver, where it is divided between the QKD receiver and the classical receiver by a dichroic mirror. From the classical receiver, the clock is recovered with a CDR and locked into a phase-locking loop (PLL) that generates the reference signal for the QKD receiver.
The QKD receiver is composed of a polarization analyzer, followed by four silicon single-photon detectors, and a time-to-digital converter (TDC) with $81ps$ resolution.
Each photodetector is measuring a different polarization state: $|H\rangle$ horizontal, $|V\rangle$ vertical, $| D\rangle$ diagonal, and $|A\rangle$ antidiagonal.

For the classical communication channel, a $1.25 Gbit$ on-off keying (OOK) signal has been implemented by a Xilinx Kintex UltraScale FPGA (KCU105E) evaluation board, an FMC connectivity board for the transceiver interface (wearing the low-jitter $156.25 MHz$ Clock source), and an optical transceiver module supporting up to $11.3 Gbps$. The module transmitter consists of an electro-absorption modulated laser (EML) ($1552.52 nm$ wavelength) and its driver, the receiver of a photodetector, and an amplification and quantification block. This module has an output power of $0 dBm$ and a $10^{-12}$ BER sensitivity of $-20 dBm$ sensitivity for the used speed. An SMF connects the optical module to the collimators.
The IBERT Ultrascale GTH (1.4) from Xilinx is used to generate a PRBS 31-bit pattern, which then sends and receives the data stream over the GTH transceivers to and from the optical module. 
The RX clock data recovery (CDR) circuit of the GTH transceiver extracts the recovered clock and data from the incoming data stream employing a phase rotator architecture ~\cite{xilGTH}. Additional PLL blocks are used to convert and provide clocks to the QKD source ($10 MHz$) and to the QKD receiver ($10 kHz$) from the recovered clock on the communication receiver. The TX-to-RX jitter provided to the QKD blocks has been measured to be smaller than $0.1 ns$. The jitters of the clocks themselves were smaller than $0.03 ns$.

\section{Results}
\begin{figure}[!h]
\begin{center}
  \includegraphics[width=0.5\linewidth]{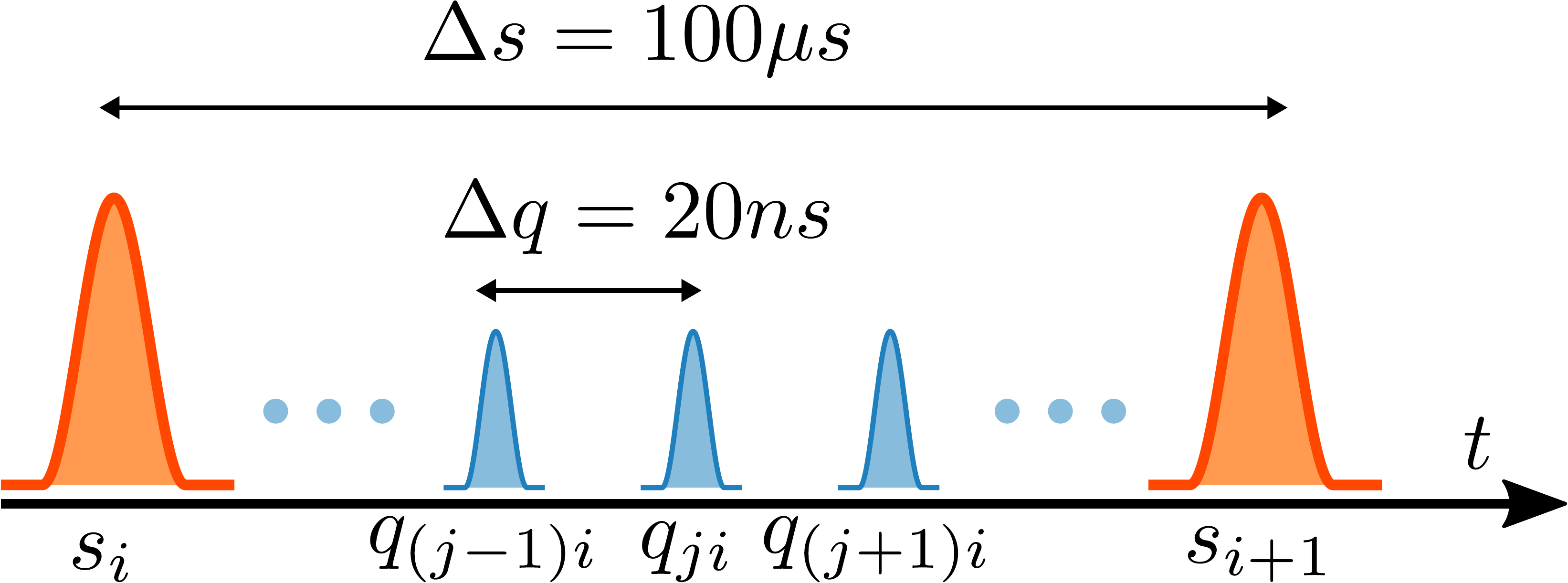}
  \caption{Temporal relation between synchronization pulses and qubits used in our experiment. The taller orange pulses represent the electrical pulse signal generated by the classical receiver after the CDR procedure, whereas the lower blue pulses represent the quantum signals detected by the QKD receiver.
  }
  \label{fig:peaks-synch}
\end{center}
\end{figure}
The performances of the new method used for clock recovery are evaluated by comparing the two setups shown in Fig.\ref{fig:SetupsSchemes}. Three experiments are performed: recovery of the qubit distribution, decimation of the reference signal at the receiver, and quantum bit error rate (QBER) estimation in a simulated communication.

\begin{figure}[!h]
  \includegraphics[width=\linewidth]{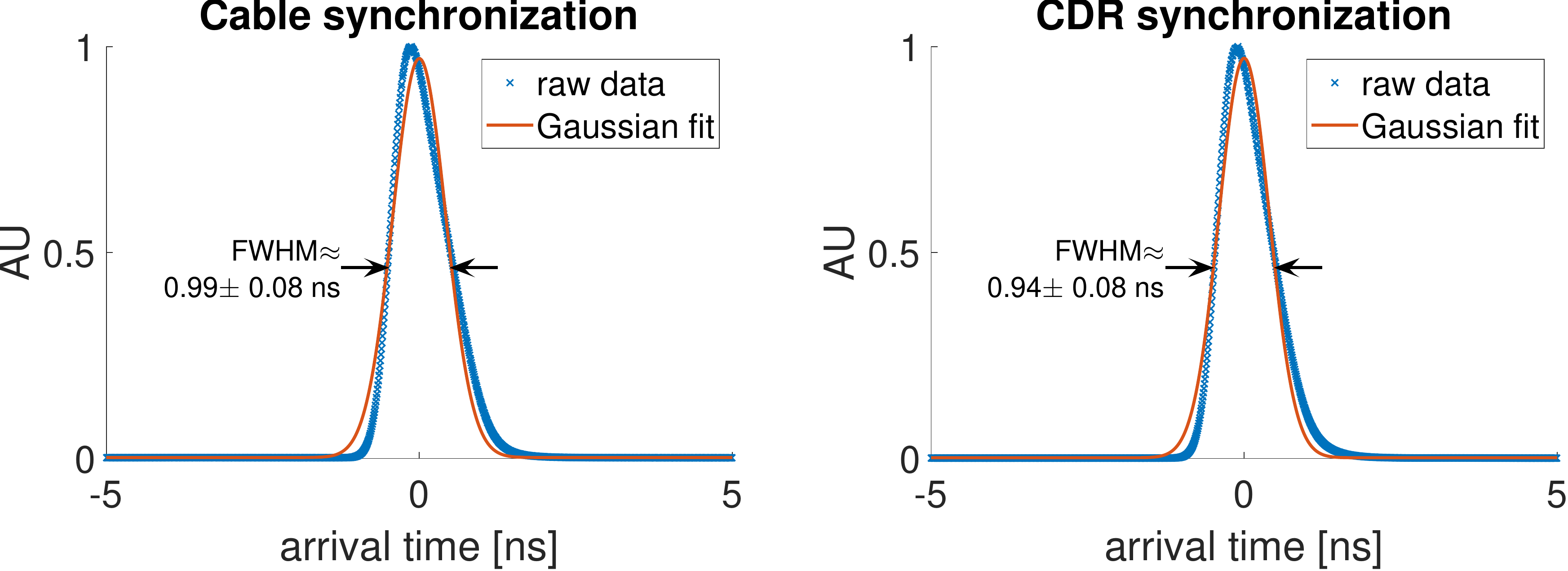}
  \caption{Side-by-side comparison of the arrival time distribution obtained by direct cable synchronization (left) and by the clock-data-recovery routine on the classical signal (right). The crosses are the measured values obtained from the setup, while the continuous line is the Gaussian fit.}
  \label{fig:fwhmComparison}
\end{figure}

The first experiment aims to recover the qubit arrival distribution. It is obtained by counting the number of photons arriving in a qubit slot interval of $20 ns$ using $15$ minutes of integration time. The measured arrival times $q_{ji}$ are re-scaled into $q'_{ji}$ according to the $10 kHz$ reference signal using:
\begin{equation}
    q'_{ji} = \frac{q_{ji} - s_i}{s_{i+1} - s_i} \cdot \Delta s
    \label{eq:QubitRescale}
\end{equation}
where $s_i$ is the measured synchronization pulse before the qubit, $s_{i+1}$ is the synchronization interval adjacent on the right of the qubits, and $\Delta s = 100 \mu s$ is the theoretical synchronization slot size when using a $10 kHz$ reference signal (see Fig.\ref{fig:peaks-synch}).
The plots in Fig.\ref{fig:fwhmComparison} show the histogram of the difference between the rescaled detections and the expected arrival times obtained by:
\begin{equation}
    a_i = \textrm{mod}(q'_{ji}, \Delta q)
\end{equation}
where $\Delta q = 20 ns$ is the theoretical qubit slot size.
The two peaks obtained by synchronization with classical communication (left) and by direct cable connection (right) have comparable shapes and full-width half-maximum (FWHM). The two FWHM are obtained by fitting the histograms with a Gaussian fit.

This demonstrates that the performance of the CDR synchronization procedure is fully equivalent to the performance obtained with cable synchronization, and the expected time of arrival can be recovered precisely. The residual error of about $1 ns$ is mainly due to the intrinsic jitter of the used single photon detectors and the TDC. It is worth noting that our scheme allows sharing the clock frequency between the transmitter and the receiver, while the phase of the signal (corresponding to the time of arrival of the first qubit) is recovered in post-processing.
Indeed, the CDR procedure only conveys frequency information, whereas both frequency and absolute time information are required for full synchronization.
In order to perform full synchronization in real-time, the classical communication protocol would
need to be adapted to transmit absolute time information at regular intervals.

\begin{figure}[!h]
  \includegraphics[width=\linewidth]{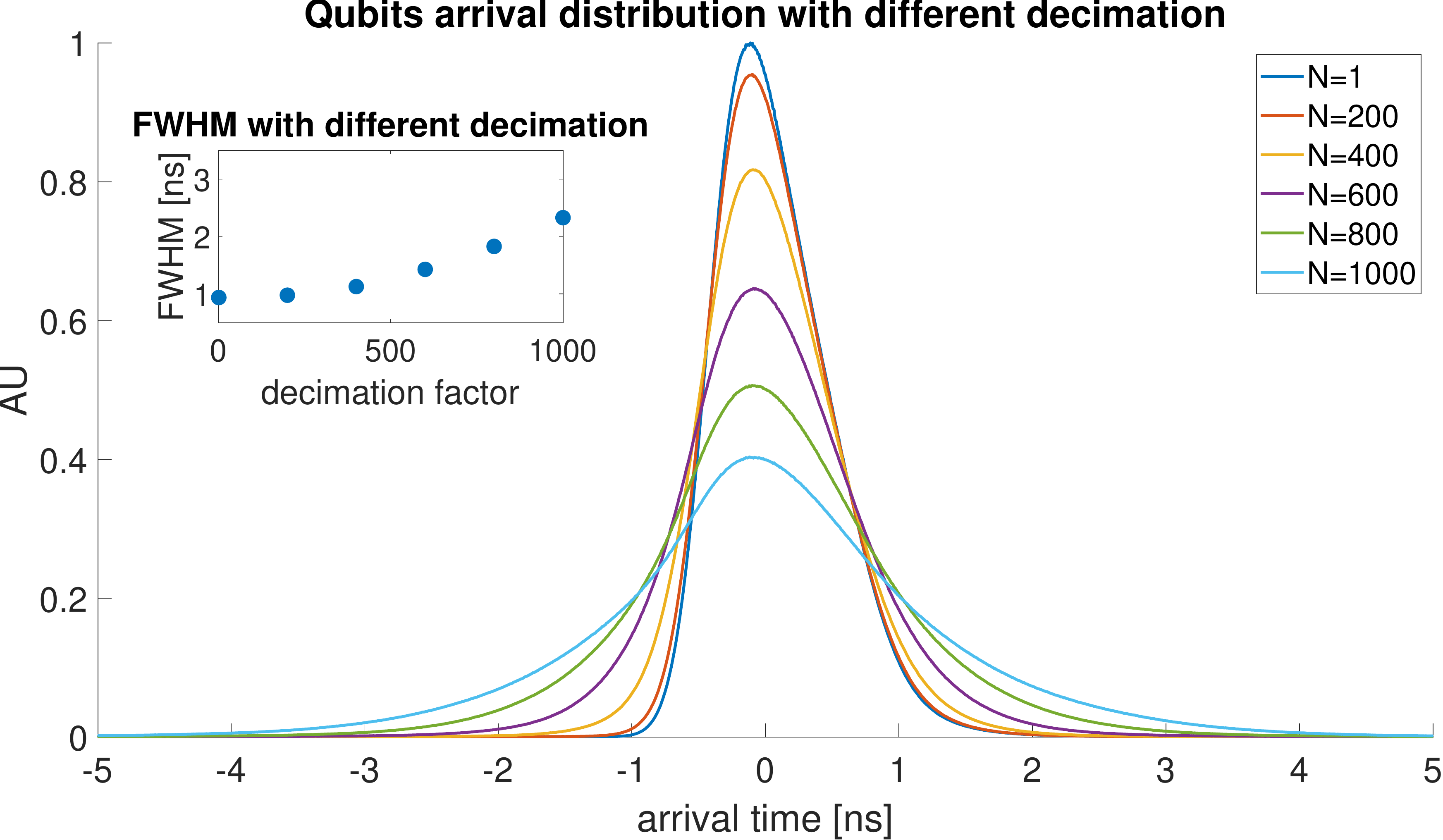}
  \caption{The figure shows the arrival distribution of each peak making varying the decimation of the synchronization. In the inset, we show the FWHM trend. Each FWHM is obtained by a Gaussian fit of the arrival distributions for a given decimation factor. 
  }
  \label{fig:decimations}
\end{figure}

The second experiment applies the same analysis as before but uses a decimated reference signal obtained by decimating the $10 kHz$ signal with different values $N$, simulating a larger interval between synchronization pulses $\Delta s = N \times 100\mu s $ that implies a lower bandwidth for the clock recovery. As shown in Fig.\ref{fig:decimations}, the peaks begin to spread while increasing the decimation factor. This behavior is expected since by increasing the interval between reference pulses, the frequency difference between the transmitter and receiver clocks integrates over a longer period making the correction factor in Eq.~\ref{eq:QubitRescale} less precise. It is noticeable that after $N=200$ (i.e., $50 Hz$ and $\Delta s = 20 ms$) the FWHM increases drastically, suggesting that a good limit for the reference clock should be above $50 Hz$. This result depends on the relative stability between the source clock and the TDC clock used in this experiment.

\begin{figure}[!h]
  \includegraphics[width=\linewidth]{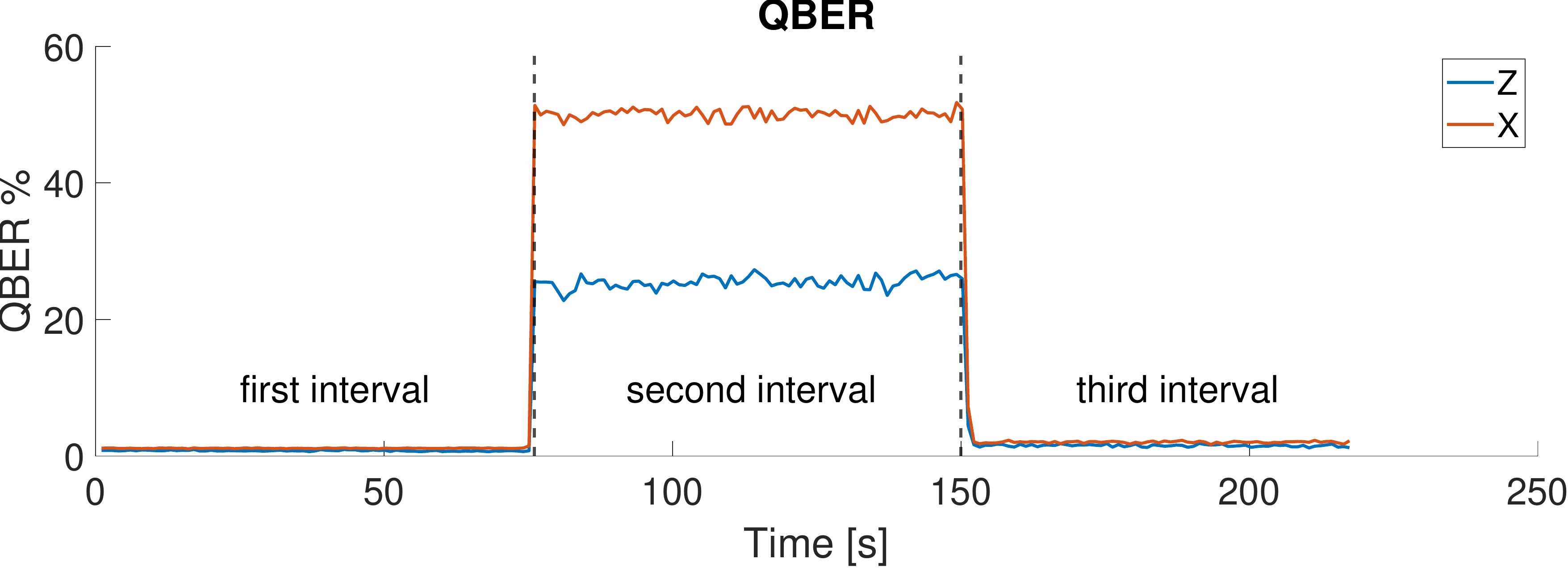}
  \caption{The QBER of the communication over time when the classical signal is blocked. At the 75th second, the classical channel is blocked and the system loses synchronization, increasing the QBER. At the 150th second, the classical channel is unlocked and the synchronization is recovered.}
  \label{fig:qber}
\end{figure}

The third test was performed to explore what will happen if the synchronization signal is interrupted. It is divided into three temporal intervals, each one minute long: the first minute of QKD with synchronization active, the second minute in which the classical channel is blocked by placing a beam dump before the dichroic mirror at the source, and the third minute where the classical channel is unblocked. The QKD communication was performed with 25\% $\ket{H}$, 25\% $\ket{V}$, and 50\% $\ket{D}  = \left( \ket{H} + \ket{V} \right ) / \sqrt{2}$ states.
As shown in Fig.\ref{fig:qber}, when the optical signal is blocked (second interval), the $10 kHz$ signal at the receiver is no longer synchronized with the source clock. Therefore, the receiver is no longer able to predict the arrival time of the qubits, and the QBER increases. This is equivalent to associating with each detection one state randomly between H, V, and D. Due to the fact that the D state is sent twice more often than the H and V states, QBER is expected at 50\% in the $Z=\{\ket{H},\ket{V} \}$ basis and 25\% in the $X=\{\ket{D}, \ket{A} = \left( \ket{H} - \ket{V} \right ) / \sqrt{2} \}$ basis. In both the first and third intervals, some post-processing was used to identify the time-of-arrival of the first qubit of the considered interval.
It is worth noting that because both quantum and classical signals are co-propagating, it is unlikable to shadow only the classical signal which is by design a stronger signal. For this reason, this experiment is not trying to reproduce a physical situation but to show that the synchronization information of the quantum channel is carried by the classical one.

\section{Conclusion}
After the experiments conducted in the laboratory, the performance of our technique \ref{fig:SetupsSchemes}-b measured in FWHM (shown in Fig.\ref{fig:fwhmComparison}) is comparable to the connection of typical test environments \ref{fig:SetupsSchemes}-a. This type of synchronization also shows in Fig.\ref{fig:decimations} some versatility in choosing synchronization repetition rates, allowing the matching of receivers with different clock requirements. Future research can be done to test this strategy in real-world environments, such as a 20 km free-space link to study the impact of atmospheric phenomena such as turbulence~\cite{Scriminich2022}. Regarding the use of this technique in satellite-based links, since both classical and quantum signals are co-propagating, the Doppler effect affects both classical and quantum signals in the same way, and therefore no additional processing would be required after the CDR: the clock recovered by the CDR will be locked to the qubit stream since they both undergo the same Doppler shift with respect to the source clock.

This type of synchronization could be really beneficial for fiber communication, where classical communication and quantum communication share the same link. In this case, the environment is more stable and the classical communication is more convenient to be optical instead of RF.

As mentioned above, future works will consist of the demonstration of full synchronization in real-time, by adapting the classical communication protocol to transmit absolute time information at regular intervals. Such time information represents the start of a qubit frame. The length of such frames will depend on the availability of the classical link: if the classical link is stable (as it happens with fiber links) over the full session, a single frame is sufficient. 
If the classical link can be interrupted (as it happens with satellite connections), the synchronization will be lost until the start of the subsequent frame.

\begin{acknowledgments}

F.B., C.A., A.S., M.A., P.V., and G.V. developed the QKD system. M.K. and D.M. developed the classical communication system. F.B., C.A., M.K., and D.M. performed and analyzed the experiment. All authors discussed the results. F.B. wrote the manuscript with inputs from all the authors.

We thank Sandrine M\"uller for administrative support.
{Funded by the EU. ESA has received funds in its quality as funding body under the European Union’s Horizon 2020 research and innovation programme}
H2020-ESA-038.12 GNSS EVOLUTIONS EXPERIMENTAL PAYLOADS AND SCIENCE ACTIVITIES - CALL FOR IDEAS
[ESA contract number:
4000130788/20/NL/AS] 

The view expressed herein can in no way be taken to reflect the official opinion of the European Union and/or the European Space Agency. Neither the European Union nor the European Space Agency shall be responsible for any use that may be made of the information it contains.
\end{acknowledgments}

\bibliographystyle{apsrev4-2}
\bibliography{synchronizationOverClassicalChannel}

\providecommand{\noopsort}[1]{}\providecommand{\singleletter}[1]{#1}%
\begin{thebibliography}{24}%
\makeatletter
\providecommand \@ifxundefined [1]{%
 \@ifx{#1\undefined}
}%
\providecommand \@ifnum [1]{%
 \ifnum #1\expandafter \@firstoftwo
 \else \expandafter \@secondoftwo
 \fi
}%
\providecommand \@ifx [1]{%
 \ifx #1\expandafter \@firstoftwo
 \else \expandafter \@secondoftwo
 \fi
}%
\providecommand \natexlab [1]{#1}%
\providecommand \enquote  [1]{``#1''}%
\providecommand \bibnamefont  [1]{#1}%
\providecommand \bibfnamefont [1]{#1}%
\providecommand \citenamefont [1]{#1}%
\providecommand \href@noop [0]{\@secondoftwo}%
\providecommand \href [0]{\begingroup \@sanitize@url \@href}%
\providecommand \@href[1]{\@@startlink{#1}\@@href}%
\providecommand \@@href[1]{\endgroup#1\@@endlink}%
\providecommand \@sanitize@url [0]{\catcode `\\12\catcode `\$12\catcode
  `\&12\catcode `\#12\catcode `\^12\catcode `\_12\catcode `\%12\relax}%
\providecommand \@@startlink[1]{}%
\providecommand \@@endlink[0]{}%
\providecommand \url  [0]{\begingroup\@sanitize@url \@url }%
\providecommand \@url [1]{\endgroup\@href {#1}{\urlprefix }}%
\providecommand \urlprefix  [0]{URL }%
\providecommand \Eprint [0]{\href }%
\providecommand \doibase [0]{https://doi.org/}%
\providecommand \selectlanguage [0]{\@gobble}%
\providecommand \bibinfo  [0]{\@secondoftwo}%
\providecommand \bibfield  [0]{\@secondoftwo}%
\providecommand \translation [1]{[#1]}%
\providecommand \BibitemOpen [0]{}%
\providecommand \bibitemStop [0]{}%
\providecommand \bibitemNoStop [0]{.\EOS\space}%
\providecommand \EOS [0]{\spacefactor3000\relax}%
\providecommand \BibitemShut  [1]{\csname bibitem#1\endcsname}%
\let\auto@bib@innerbib\@empty
\bibitem [{\citenamefont {Fields}\ \emph {et~al.}(2009)\citenamefont {Fields},
  \citenamefont {Lunde}, \citenamefont {Wong}, \citenamefont {Wicker},
  \citenamefont {Kozlowski}, \citenamefont {Jordan}, \citenamefont {Hansen},
  \citenamefont {Muehlnikel}, \citenamefont {Scheel}, \citenamefont {Sterr},
  \citenamefont {Kahle},\ and\ \citenamefont {Meyer}}]{Fields2009}%
  \BibitemOpen
  \bibfield  {author} {\bibinfo {author} {\bibfnamefont {R.}~\bibnamefont
  {Fields}}, \bibinfo {author} {\bibfnamefont {C.}~\bibnamefont {Lunde}},
  \bibinfo {author} {\bibfnamefont {R.}~\bibnamefont {Wong}}, \bibinfo {author}
  {\bibfnamefont {J.}~\bibnamefont {Wicker}}, \bibinfo {author} {\bibfnamefont
  {D.}~\bibnamefont {Kozlowski}}, \bibinfo {author} {\bibfnamefont
  {J.}~\bibnamefont {Jordan}}, \bibinfo {author} {\bibfnamefont
  {B.}~\bibnamefont {Hansen}}, \bibinfo {author} {\bibfnamefont
  {G.}~\bibnamefont {Muehlnikel}}, \bibinfo {author} {\bibfnamefont
  {W.}~\bibnamefont {Scheel}}, \bibinfo {author} {\bibfnamefont
  {U.}~\bibnamefont {Sterr}}, \bibinfo {author} {\bibfnamefont
  {R.}~\bibnamefont {Kahle}},\ and\ \bibinfo {author} {\bibfnamefont
  {R.}~\bibnamefont {Meyer}},\ }in\ \href {https://doi.org/10.1117/12.820393}
  {\emph {\bibinfo {booktitle} {Sensors and Systems for Space Applications
  III}}},\ Vol.\ \bibinfo {volume} {7330},\ \bibinfo {editor} {edited by\
  \bibinfo {editor} {\bibfnamefont {J.~L.}\ \bibnamefont {Cox}}\ and\ \bibinfo
  {editor} {\bibfnamefont {P.}~\bibnamefont {Motaghedi}}},\ \bibinfo
  {organization} {International Society for Optics and Photonics}\ (\bibinfo
  {publisher} {SPIE},\ \bibinfo {year} {2009})\ pp.\ \bibinfo {pages} {211 --
  225}\BibitemShut {NoStop}%
\bibitem [{\citenamefont {Gregory}\ \emph {et~al.}(2013)\citenamefont
  {Gregory}, \citenamefont {Troendle}, \citenamefont {Muehlnikel},
  \citenamefont {Heine}, \citenamefont {Meyer}, \citenamefont {Lutzer},\ and\
  \citenamefont {Czichy}}]{Gregory2013}%
  \BibitemOpen
  \bibfield  {author} {\bibinfo {author} {\bibfnamefont {M.}~\bibnamefont
  {Gregory}}, \bibinfo {author} {\bibfnamefont {D.}~\bibnamefont {Troendle}},
  \bibinfo {author} {\bibfnamefont {G.}~\bibnamefont {Muehlnikel}}, \bibinfo
  {author} {\bibfnamefont {F.}~\bibnamefont {Heine}}, \bibinfo {author}
  {\bibfnamefont {R.}~\bibnamefont {Meyer}}, \bibinfo {author} {\bibfnamefont
  {M.}~\bibnamefont {Lutzer}},\ and\ \bibinfo {author} {\bibfnamefont
  {R.}~\bibnamefont {Czichy}},\ }in\ \href {https://doi.org/10.1117/12.2022253}
  {\emph {\bibinfo {booktitle} {Free-Space Laser Communication and
  Atmospheric\\ Propagation XXV}}},\ Vol.\ \bibinfo {volume} {8610},\ \bibinfo
  {editor} {edited by\ \bibinfo {editor} {\bibfnamefont {H.}~\bibnamefont
  {Hemmati}}\ and\ \bibinfo {editor} {\bibfnamefont {D.~M.}\ \bibnamefont
  {Boroson}}},\ \bibinfo {organization} {International Society for Optics and
  Photonics}\ (\bibinfo  {publisher} {SPIE},\ \bibinfo {year} {2013})\ pp.\
  \bibinfo {pages} {17 -- 29}\BibitemShut {NoStop}%
\bibitem [{\citenamefont {Carrasco-Casado}\ \emph {et~al.}(2017)\citenamefont
  {Carrasco-Casado}, \citenamefont {Takenaka}, \citenamefont {Kolev},
  \citenamefont {Munemasa}, \citenamefont {Kunimori}, \citenamefont {Suzuki},
  \citenamefont {Fuse}, \citenamefont {Kubo-Oka}, \citenamefont {Akioka},
  \citenamefont {Koyama},\ and\ \citenamefont {Toyoshima}}]{Casado2017}%
  \BibitemOpen
  \bibfield  {author} {\bibinfo {author} {\bibfnamefont {A.}~\bibnamefont
  {Carrasco-Casado}}, \bibinfo {author} {\bibfnamefont {H.}~\bibnamefont
  {Takenaka}}, \bibinfo {author} {\bibfnamefont {D.}~\bibnamefont {Kolev}},
  \bibinfo {author} {\bibfnamefont {Y.}~\bibnamefont {Munemasa}}, \bibinfo
  {author} {\bibfnamefont {H.}~\bibnamefont {Kunimori}}, \bibinfo {author}
  {\bibfnamefont {K.}~\bibnamefont {Suzuki}}, \bibinfo {author} {\bibfnamefont
  {T.}~\bibnamefont {Fuse}}, \bibinfo {author} {\bibfnamefont {T.}~\bibnamefont
  {Kubo-Oka}}, \bibinfo {author} {\bibfnamefont {M.}~\bibnamefont {Akioka}},
  \bibinfo {author} {\bibfnamefont {Y.}~\bibnamefont {Koyama}},\ and\ \bibinfo
  {author} {\bibfnamefont {M.}~\bibnamefont {Toyoshima}},\ }\href
  {https://doi.org/https://doi.org/10.1016/j.actaastro.2017.07.030} {\bibfield
  {journal} {\bibinfo  {journal} {Acta Astronautica}\ }\textbf {\bibinfo
  {volume} {139}},\ \bibinfo {pages} {377} (\bibinfo {year}
  {2017})}\BibitemShut {NoStop}%
\bibitem [{\citenamefont {Zech}\ \emph {et~al.}(2015)\citenamefont {Zech},
  \citenamefont {Heine}, \citenamefont {Tröndle}, \citenamefont {Seel},
  \citenamefont {Motzigemba}, \citenamefont {Meyer},\ and\ \citenamefont
  {Philipp-May}}]{Zech2015}%
  \BibitemOpen
  \bibfield  {author} {\bibinfo {author} {\bibfnamefont {H.}~\bibnamefont
  {Zech}}, \bibinfo {author} {\bibfnamefont {F.}~\bibnamefont {Heine}},
  \bibinfo {author} {\bibfnamefont {D.}~\bibnamefont {Tröndle}}, \bibinfo
  {author} {\bibfnamefont {S.}~\bibnamefont {Seel}}, \bibinfo {author}
  {\bibfnamefont {M.}~\bibnamefont {Motzigemba}}, \bibinfo {author}
  {\bibfnamefont {R.}~\bibnamefont {Meyer}},\ and\ \bibinfo {author}
  {\bibfnamefont {S.}~\bibnamefont {Philipp-May}},\ }in\ \href
  {https://doi.org/10.1117/12.2196273} {\emph {\bibinfo {booktitle}
  {Unmanned/Unattended Sensors and Sensor Networks XI; and Advanced Free-Space
  Optical Communication Techniques and Applications}}},\ Vol.\ \bibinfo
  {volume} {9647},\ \bibinfo {editor} {edited by\ \bibinfo {editor}
  {\bibfnamefont {E.~M.}\ \bibnamefont {Carapezza}}, \bibinfo {editor}
  {\bibfnamefont {P.~G.}\ \bibnamefont {Datskos}}, \bibinfo {editor}
  {\bibfnamefont {C.}~\bibnamefont {Tsamis}}, \bibinfo {editor} {\bibfnamefont
  {L.}~\bibnamefont {Laycock}},\ and\ \bibinfo {editor} {\bibfnamefont {H.~J.}\
  \bibnamefont {White}}},\ \bibinfo {organization} {International Society for
  Optics and Photonics}\ (\bibinfo  {publisher} {SPIE},\ \bibinfo {year}
  {2015})\ pp.\ \bibinfo {pages} {85 -- 92}\BibitemShut {NoStop}%
\bibitem [{\citenamefont {Alonso}\ \emph {et~al.}(2004)\citenamefont {Alonso},
  \citenamefont {Reyes},\ and\ \citenamefont {Sodnik}}]{Alonso2004}%
  \BibitemOpen
  \bibfield  {author} {\bibinfo {author} {\bibfnamefont {A.}~\bibnamefont
  {Alonso}}, \bibinfo {author} {\bibfnamefont {M.}~\bibnamefont {Reyes}},\ and\
  \bibinfo {author} {\bibfnamefont {Z.}~\bibnamefont {Sodnik}},\ }in\ \href
  {https://doi.org/10.1117/12.565516} {\emph {\bibinfo {booktitle} {Optics in
  Atmospheric Propagation and Adaptive Systems VII}}},\ Vol.\ \bibinfo {volume}
  {5572},\ \bibinfo {editor} {edited by\ \bibinfo {editor} {\bibfnamefont
  {J.~D.}\ \bibnamefont {Gonglewski}}\ and\ \bibinfo {editor} {\bibfnamefont
  {K.}~\bibnamefont {Stein}}},\ \bibinfo {organization} {International Society
  for Optics and Photonics}\ (\bibinfo  {publisher} {SPIE},\ \bibinfo {year}
  {2004})\ pp.\ \bibinfo {pages} {372 -- 383}\BibitemShut {NoStop}%
\bibitem [{\citenamefont {Boroson}\ \emph {et~al.}(2014)\citenamefont
  {Boroson}, \citenamefont {Robinson}, \citenamefont {Murphy}, \citenamefont
  {Burianek}, \citenamefont {Khatri}, \citenamefont {Kovalik}, \citenamefont
  {Sodnik},\ and\ \citenamefont {Cornwell}}]{Borson2014}%
  \BibitemOpen
  \bibfield  {author} {\bibinfo {author} {\bibfnamefont {D.~M.}\ \bibnamefont
  {Boroson}}, \bibinfo {author} {\bibfnamefont {B.~S.}\ \bibnamefont
  {Robinson}}, \bibinfo {author} {\bibfnamefont {D.~V.}\ \bibnamefont
  {Murphy}}, \bibinfo {author} {\bibfnamefont {D.~A.}\ \bibnamefont
  {Burianek}}, \bibinfo {author} {\bibfnamefont {F.}~\bibnamefont {Khatri}},
  \bibinfo {author} {\bibfnamefont {J.~M.}\ \bibnamefont {Kovalik}}, \bibinfo
  {author} {\bibfnamefont {Z.}~\bibnamefont {Sodnik}},\ and\ \bibinfo {author}
  {\bibfnamefont {D.~M.}\ \bibnamefont {Cornwell}},\ }in\ \href
  {https://doi.org/10.1117/12.2045508} {\emph {\bibinfo {booktitle} {Free-Space
  Laser Communication and Atmospheric Propagation XXVI}}},\ Vol.\ \bibinfo
  {volume} {8971},\ \bibinfo {editor} {edited by\ \bibinfo {editor}
  {\bibfnamefont {H.}~\bibnamefont {Hemmati}}\ and\ \bibinfo {editor}
  {\bibfnamefont {D.~M.}\ \bibnamefont {Boroson}}},\ \bibinfo {organization}
  {International Society for Optics and Photonics}\ (\bibinfo  {publisher}
  {SPIE},\ \bibinfo {year} {2014})\ pp.\ \bibinfo {pages} {213 --
  223}\BibitemShut {NoStop}%
\bibitem [{\citenamefont {Sansone}\ \emph {et~al.}(2020)\citenamefont
  {Sansone}, \citenamefont {Francesconi}, \citenamefont {Corvaja},
  \citenamefont {Vallone}, \citenamefont {Antonello}, \citenamefont {Branz},\
  and\ \citenamefont {Villoresi}}]{Sansone2020}%
  \BibitemOpen
  \bibfield  {author} {\bibinfo {author} {\bibfnamefont {F.}~\bibnamefont
  {Sansone}}, \bibinfo {author} {\bibfnamefont {A.}~\bibnamefont
  {Francesconi}}, \bibinfo {author} {\bibfnamefont {R.}~\bibnamefont
  {Corvaja}}, \bibinfo {author} {\bibfnamefont {G.}~\bibnamefont {Vallone}},
  \bibinfo {author} {\bibfnamefont {R.}~\bibnamefont {Antonello}}, \bibinfo
  {author} {\bibfnamefont {F.}~\bibnamefont {Branz}},\ and\ \bibinfo {author}
  {\bibfnamefont {P.}~\bibnamefont {Villoresi}},\ }\href
  {https://doi.org/https://doi.org/10.1016/j.actaastro.2020.04.049} {\bibfield
  {journal} {\bibinfo  {journal} {Acta Astronautica}\ }\textbf {\bibinfo
  {volume} {173}},\ \bibinfo {pages} {310} (\bibinfo {year}
  {2020})}\BibitemShut {NoStop}%
\bibitem [{\citenamefont {Pirandola}\ \emph {et~al.}(2020)\citenamefont
  {Pirandola}, \citenamefont {Andersen}, \citenamefont {Banchi}, \citenamefont
  {Berta}, \citenamefont {Bunandar}, \citenamefont {Colbeck}, \citenamefont
  {Englund}, \citenamefont {Gehring}, \citenamefont {Lupo}, \citenamefont
  {Ottaviani}, \citenamefont {Pereira}, \citenamefont {Razavi}, \citenamefont
  {{Shamsul Shaari}}, \citenamefont {Tomamichel}, \citenamefont {Usenko},
  \citenamefont {Vallone}, \citenamefont {Villoresi},\ and\ \citenamefont
  {Wallden}}]{Pirandola2020}%
  \BibitemOpen
  \bibfield  {author} {\bibinfo {author} {\bibfnamefont {S.}~\bibnamefont
  {Pirandola}}, \bibinfo {author} {\bibfnamefont {U.~L.}\ \bibnamefont
  {Andersen}}, \bibinfo {author} {\bibfnamefont {L.}~\bibnamefont {Banchi}},
  \bibinfo {author} {\bibfnamefont {M.}~\bibnamefont {Berta}}, \bibinfo
  {author} {\bibfnamefont {D.}~\bibnamefont {Bunandar}}, \bibinfo {author}
  {\bibfnamefont {R.}~\bibnamefont {Colbeck}}, \bibinfo {author} {\bibfnamefont
  {D.}~\bibnamefont {Englund}}, \bibinfo {author} {\bibfnamefont
  {T.}~\bibnamefont {Gehring}}, \bibinfo {author} {\bibfnamefont
  {C.}~\bibnamefont {Lupo}}, \bibinfo {author} {\bibfnamefont {C.}~\bibnamefont
  {Ottaviani}}, \bibinfo {author} {\bibfnamefont {J.~L.}\ \bibnamefont
  {Pereira}}, \bibinfo {author} {\bibfnamefont {M.}~\bibnamefont {Razavi}},
  \bibinfo {author} {\bibfnamefont {J.}~\bibnamefont {{Shamsul Shaari}}},
  \bibinfo {author} {\bibfnamefont {M.}~\bibnamefont {Tomamichel}}, \bibinfo
  {author} {\bibfnamefont {V.~C.}\ \bibnamefont {Usenko}}, \bibinfo {author}
  {\bibfnamefont {G.}~\bibnamefont {Vallone}}, \bibinfo {author} {\bibfnamefont
  {P.}~\bibnamefont {Villoresi}},\ and\ \bibinfo {author} {\bibfnamefont
  {P.}~\bibnamefont {Wallden}},\ }\href {https://doi.org/10.1364/AOP.361502}
  {\bibfield  {journal} {\bibinfo  {journal} {Adv. Opt. Photonics}\ }\textbf
  {\bibinfo {volume} {12}},\ \bibinfo {pages} {1012} (\bibinfo {year}
  {2020})}\BibitemShut {NoStop}%
\bibitem [{\citenamefont {Scarani}\ \emph {et~al.}(2009)\citenamefont
  {Scarani}, \citenamefont {Bechmann-Pasquinucci}, \citenamefont {Cerf},
  \citenamefont {Du\ifmmode~\check{s}\else \v{s}\fi{}ek}, \citenamefont
  {L\"utkenhaus},\ and\ \citenamefont {Peev}}]{Scarani2008}%
  \BibitemOpen
  \bibfield  {author} {\bibinfo {author} {\bibfnamefont {V.}~\bibnamefont
  {Scarani}}, \bibinfo {author} {\bibfnamefont {H.}~\bibnamefont
  {Bechmann-Pasquinucci}}, \bibinfo {author} {\bibfnamefont {N.~J.}\
  \bibnamefont {Cerf}}, \bibinfo {author} {\bibfnamefont {M.}~\bibnamefont
  {Du\ifmmode~\check{s}\else \v{s}\fi{}ek}}, \bibinfo {author} {\bibfnamefont
  {N.}~\bibnamefont {L\"utkenhaus}},\ and\ \bibinfo {author} {\bibfnamefont
  {M.}~\bibnamefont {Peev}},\ }\href
  {https://doi.org/10.1103/RevModPhys.81.1301} {\bibfield  {journal} {\bibinfo
  {journal} {Rev. Mod. Phys.}\ }\textbf {\bibinfo {volume} {81}},\ \bibinfo
  {pages} {1301} (\bibinfo {year} {2009})}\BibitemShut {NoStop}%
\bibitem [{\citenamefont {Boaron}\ \emph {et~al.}(2018)\citenamefont {Boaron},
  \citenamefont {Boso}, \citenamefont {Rusca}, \citenamefont {Vulliez},
  \citenamefont {Autebert}, \citenamefont {Caloz}, \citenamefont {Perrenoud},
  \citenamefont {Gras}, \citenamefont {Bussi\`eres}, \citenamefont {Li},
  \citenamefont {Nolan}, \citenamefont {Martin},\ and\ \citenamefont
  {Zbinden}}]{Boaron2018}%
  \BibitemOpen
  \bibfield  {author} {\bibinfo {author} {\bibfnamefont {A.}~\bibnamefont
  {Boaron}}, \bibinfo {author} {\bibfnamefont {G.}~\bibnamefont {Boso}},
  \bibinfo {author} {\bibfnamefont {D.}~\bibnamefont {Rusca}}, \bibinfo
  {author} {\bibfnamefont {C.}~\bibnamefont {Vulliez}}, \bibinfo {author}
  {\bibfnamefont {C.}~\bibnamefont {Autebert}}, \bibinfo {author}
  {\bibfnamefont {M.}~\bibnamefont {Caloz}}, \bibinfo {author} {\bibfnamefont
  {M.}~\bibnamefont {Perrenoud}}, \bibinfo {author} {\bibfnamefont
  {G.}~\bibnamefont {Gras}}, \bibinfo {author} {\bibfnamefont {F.}~\bibnamefont
  {Bussi\`eres}}, \bibinfo {author} {\bibfnamefont {M.-J.}\ \bibnamefont {Li}},
  \bibinfo {author} {\bibfnamefont {D.}~\bibnamefont {Nolan}}, \bibinfo
  {author} {\bibfnamefont {A.}~\bibnamefont {Martin}},\ and\ \bibinfo {author}
  {\bibfnamefont {H.}~\bibnamefont {Zbinden}},\ }\href
  {https://doi.org/10.1103/PhysRevLett.121.190502} {\bibfield  {journal}
  {\bibinfo  {journal} {Phys. Rev. Lett.}\ }\textbf {\bibinfo {volume} {121}},\
  \bibinfo {pages} {190502} (\bibinfo {year} {2018})}\BibitemShut {NoStop}%
\bibitem [{\citenamefont {Liao}\ \emph {et~al.}(2017)\citenamefont {Liao},
  \citenamefont {Cai}, \citenamefont {Liu}, \citenamefont {Zhang},
  \citenamefont {Li}, \citenamefont {Ren}, \citenamefont {Yin}, \citenamefont
  {Shen}, \citenamefont {Cao}, \citenamefont {Li}, \citenamefont {Li},
  \citenamefont {Chen}, \citenamefont {Sun}, \citenamefont {Jia}, \citenamefont
  {Wu}, \citenamefont {Jiang}, \citenamefont {Wang}, \citenamefont {Huang},
  \citenamefont {Wang}, \citenamefont {Zhou}, \citenamefont {Deng},
  \citenamefont {Xi}, \citenamefont {Ma}, \citenamefont {Hu}, \citenamefont
  {Zhang}, \citenamefont {Chen}, \citenamefont {Liu}, \citenamefont {Wang},
  \citenamefont {Zhu}, \citenamefont {Lu}, \citenamefont {Shu}, \citenamefont
  {Peng}, \citenamefont {Wang},\ and\ \citenamefont {Pan}}]{Liao2017_Sat}%
  \BibitemOpen
  \bibfield  {author} {\bibinfo {author} {\bibfnamefont {S.-K.}\ \bibnamefont
  {Liao}}, \bibinfo {author} {\bibfnamefont {W.-Q.}\ \bibnamefont {Cai}},
  \bibinfo {author} {\bibfnamefont {W.-Y.}\ \bibnamefont {Liu}}, \bibinfo
  {author} {\bibfnamefont {L.}~\bibnamefont {Zhang}}, \bibinfo {author}
  {\bibfnamefont {Y.}~\bibnamefont {Li}}, \bibinfo {author} {\bibfnamefont
  {J.-G.}\ \bibnamefont {Ren}}, \bibinfo {author} {\bibfnamefont
  {J.}~\bibnamefont {Yin}}, \bibinfo {author} {\bibfnamefont {Q.}~\bibnamefont
  {Shen}}, \bibinfo {author} {\bibfnamefont {Y.}~\bibnamefont {Cao}}, \bibinfo
  {author} {\bibfnamefont {Z.-P.}\ \bibnamefont {Li}}, \bibinfo {author}
  {\bibfnamefont {F.-Z.}\ \bibnamefont {Li}}, \bibinfo {author} {\bibfnamefont
  {X.-W.}\ \bibnamefont {Chen}}, \bibinfo {author} {\bibfnamefont {L.-H.}\
  \bibnamefont {Sun}}, \bibinfo {author} {\bibfnamefont {J.-J.}\ \bibnamefont
  {Jia}}, \bibinfo {author} {\bibfnamefont {J.-C.}\ \bibnamefont {Wu}},
  \bibinfo {author} {\bibfnamefont {X.-J.}\ \bibnamefont {Jiang}}, \bibinfo
  {author} {\bibfnamefont {J.-F.}\ \bibnamefont {Wang}}, \bibinfo {author}
  {\bibfnamefont {Y.-M.}\ \bibnamefont {Huang}}, \bibinfo {author}
  {\bibfnamefont {Q.}~\bibnamefont {Wang}}, \bibinfo {author} {\bibfnamefont
  {Y.-L.}\ \bibnamefont {Zhou}}, \bibinfo {author} {\bibfnamefont
  {L.}~\bibnamefont {Deng}}, \bibinfo {author} {\bibfnamefont {T.}~\bibnamefont
  {Xi}}, \bibinfo {author} {\bibfnamefont {L.}~\bibnamefont {Ma}}, \bibinfo
  {author} {\bibfnamefont {T.}~\bibnamefont {Hu}}, \bibinfo {author}
  {\bibfnamefont {Q.}~\bibnamefont {Zhang}}, \bibinfo {author} {\bibfnamefont
  {Y.-A.}\ \bibnamefont {Chen}}, \bibinfo {author} {\bibfnamefont {N.-L.}\
  \bibnamefont {Liu}}, \bibinfo {author} {\bibfnamefont {X.-B.}\ \bibnamefont
  {Wang}}, \bibinfo {author} {\bibfnamefont {Z.-C.}\ \bibnamefont {Zhu}},
  \bibinfo {author} {\bibfnamefont {C.-Y.}\ \bibnamefont {Lu}}, \bibinfo
  {author} {\bibfnamefont {R.}~\bibnamefont {Shu}}, \bibinfo {author}
  {\bibfnamefont {C.-Z.}\ \bibnamefont {Peng}}, \bibinfo {author}
  {\bibfnamefont {J.-Y.}\ \bibnamefont {Wang}},\ and\ \bibinfo {author}
  {\bibfnamefont {J.-W.}\ \bibnamefont {Pan}},\ }\href
  {https://doi.org/10.1038/nature23655} {\bibfield  {journal} {\bibinfo
  {journal} {Nature}\ }\textbf {\bibinfo {volume} {549}},\ \bibinfo {pages}
  {43} (\bibinfo {year} {2017})}\BibitemShut {NoStop}%
\bibitem [{\citenamefont {Liao}\ \emph {et~al.}(2018)\citenamefont {Liao},
  \citenamefont {Cai}, \citenamefont {Handsteiner}, \citenamefont {Liu},
  \citenamefont {Yin}, \citenamefont {Zhang}, \citenamefont {Rauch},
  \citenamefont {Fink}, \citenamefont {Ren}, \citenamefont {Liu}, \citenamefont
  {Li}, \citenamefont {Shen}, \citenamefont {Cao}, \citenamefont {Li},
  \citenamefont {Wang}, \citenamefont {Huang}, \citenamefont {Deng},
  \citenamefont {Xi}, \citenamefont {Ma}, \citenamefont {Hu}, \citenamefont
  {Li}, \citenamefont {Liu}, \citenamefont {Koidl}, \citenamefont {Wang},
  \citenamefont {Chen}, \citenamefont {Wang}, \citenamefont {Steindorfer},
  \citenamefont {Kirchner}, \citenamefont {Lu}, \citenamefont {Shu},
  \citenamefont {Ursin}, \citenamefont {Scheidl}, \citenamefont {Peng},
  \citenamefont {Wang}, \citenamefont {Zeilinger},\ and\ \citenamefont
  {Pan}}]{Liao2018}%
  \BibitemOpen
  \bibfield  {author} {\bibinfo {author} {\bibfnamefont {S.-K.}\ \bibnamefont
  {Liao}}, \bibinfo {author} {\bibfnamefont {W.-Q.}\ \bibnamefont {Cai}},
  \bibinfo {author} {\bibfnamefont {J.}~\bibnamefont {Handsteiner}}, \bibinfo
  {author} {\bibfnamefont {B.}~\bibnamefont {Liu}}, \bibinfo {author}
  {\bibfnamefont {J.}~\bibnamefont {Yin}}, \bibinfo {author} {\bibfnamefont
  {L.}~\bibnamefont {Zhang}}, \bibinfo {author} {\bibfnamefont
  {D.}~\bibnamefont {Rauch}}, \bibinfo {author} {\bibfnamefont
  {M.}~\bibnamefont {Fink}}, \bibinfo {author} {\bibfnamefont {J.-G.}\
  \bibnamefont {Ren}}, \bibinfo {author} {\bibfnamefont {W.-Y.}\ \bibnamefont
  {Liu}}, \bibinfo {author} {\bibfnamefont {Y.}~\bibnamefont {Li}}, \bibinfo
  {author} {\bibfnamefont {Q.}~\bibnamefont {Shen}}, \bibinfo {author}
  {\bibfnamefont {Y.}~\bibnamefont {Cao}}, \bibinfo {author} {\bibfnamefont
  {F.-Z.}\ \bibnamefont {Li}}, \bibinfo {author} {\bibfnamefont {J.-F.}\
  \bibnamefont {Wang}}, \bibinfo {author} {\bibfnamefont {Y.-M.}\ \bibnamefont
  {Huang}}, \bibinfo {author} {\bibfnamefont {L.}~\bibnamefont {Deng}},
  \bibinfo {author} {\bibfnamefont {T.}~\bibnamefont {Xi}}, \bibinfo {author}
  {\bibfnamefont {L.}~\bibnamefont {Ma}}, \bibinfo {author} {\bibfnamefont
  {T.}~\bibnamefont {Hu}}, \bibinfo {author} {\bibfnamefont {L.}~\bibnamefont
  {Li}}, \bibinfo {author} {\bibfnamefont {N.-L.}\ \bibnamefont {Liu}},
  \bibinfo {author} {\bibfnamefont {F.}~\bibnamefont {Koidl}}, \bibinfo
  {author} {\bibfnamefont {P.}~\bibnamefont {Wang}}, \bibinfo {author}
  {\bibfnamefont {Y.-A.}\ \bibnamefont {Chen}}, \bibinfo {author}
  {\bibfnamefont {X.-B.}\ \bibnamefont {Wang}}, \bibinfo {author}
  {\bibfnamefont {M.}~\bibnamefont {Steindorfer}}, \bibinfo {author}
  {\bibfnamefont {G.}~\bibnamefont {Kirchner}}, \bibinfo {author}
  {\bibfnamefont {C.-Y.}\ \bibnamefont {Lu}}, \bibinfo {author} {\bibfnamefont
  {R.}~\bibnamefont {Shu}}, \bibinfo {author} {\bibfnamefont {R.}~\bibnamefont
  {Ursin}}, \bibinfo {author} {\bibfnamefont {T.}~\bibnamefont {Scheidl}},
  \bibinfo {author} {\bibfnamefont {C.-Z.}\ \bibnamefont {Peng}}, \bibinfo
  {author} {\bibfnamefont {J.-Y.}\ \bibnamefont {Wang}}, \bibinfo {author}
  {\bibfnamefont {A.}~\bibnamefont {Zeilinger}},\ and\ \bibinfo {author}
  {\bibfnamefont {J.-W.}\ \bibnamefont {Pan}},\ }\href
  {https://doi.org/10.1103/PhysRevLett.120.030501} {\bibfield  {journal}
  {\bibinfo  {journal} {Phys. Rev. Lett.}\ }\textbf {\bibinfo {volume} {120}},\
  \bibinfo {pages} {030501} (\bibinfo {year} {2018})}\BibitemShut {NoStop}%
\bibitem [{\citenamefont {Agnesi}\ \emph {et~al.}(2019)\citenamefont {Agnesi},
  \citenamefont {Calderaro}, \citenamefont {Dequal}, \citenamefont {Vedovato},
  \citenamefont {Schiavon}, \citenamefont {Santamato}, \citenamefont {Luceri},
  \citenamefont {Bianco}, \citenamefont {Vallone},\ and\ \citenamefont
  {Villoresi}}]{Agnesi:19}%
  \BibitemOpen
  \bibfield  {author} {\bibinfo {author} {\bibfnamefont {C.}~\bibnamefont
  {Agnesi}}, \bibinfo {author} {\bibfnamefont {L.}~\bibnamefont {Calderaro}},
  \bibinfo {author} {\bibfnamefont {D.}~\bibnamefont {Dequal}}, \bibinfo
  {author} {\bibfnamefont {F.}~\bibnamefont {Vedovato}}, \bibinfo {author}
  {\bibfnamefont {M.}~\bibnamefont {Schiavon}}, \bibinfo {author}
  {\bibfnamefont {A.}~\bibnamefont {Santamato}}, \bibinfo {author}
  {\bibfnamefont {V.}~\bibnamefont {Luceri}}, \bibinfo {author} {\bibfnamefont
  {G.}~\bibnamefont {Bianco}}, \bibinfo {author} {\bibfnamefont
  {G.}~\bibnamefont {Vallone}},\ and\ \bibinfo {author} {\bibfnamefont
  {P.}~\bibnamefont {Villoresi}},\ }\href
  {https://doi.org/10.1364/JOSAB.36.000B59} {\bibfield  {journal} {\bibinfo
  {journal} {J. Opt. Soc. Am. B}\ }\textbf {\bibinfo {volume} {36}},\ \bibinfo
  {pages} {B59} (\bibinfo {year} {2019})}\BibitemShut {NoStop}%
\bibitem [{\citenamefont {Calderaro}\ \emph {et~al.}(2020)\citenamefont
  {Calderaro}, \citenamefont {Stanco}, \citenamefont {Agnesi}, \citenamefont
  {Avesani}, \citenamefont {Dequal}, \citenamefont {Villoresi},\ and\
  \citenamefont {Vallone}}]{Calderaro2020}%
  \BibitemOpen
  \bibfield  {author} {\bibinfo {author} {\bibfnamefont {L.}~\bibnamefont
  {Calderaro}}, \bibinfo {author} {\bibfnamefont {A.}~\bibnamefont {Stanco}},
  \bibinfo {author} {\bibfnamefont {C.}~\bibnamefont {Agnesi}}, \bibinfo
  {author} {\bibfnamefont {M.}~\bibnamefont {Avesani}}, \bibinfo {author}
  {\bibfnamefont {D.}~\bibnamefont {Dequal}}, \bibinfo {author} {\bibfnamefont
  {P.}~\bibnamefont {Villoresi}},\ and\ \bibinfo {author} {\bibfnamefont
  {G.}~\bibnamefont {Vallone}},\ }\href
  {https://doi.org/10.1103/PhysRevApplied.13.054041} {\bibfield  {journal}
  {\bibinfo  {journal} {Phys. Rev. Applied}\ }\textbf {\bibinfo {volume}
  {13}},\ \bibinfo {pages} {054041} (\bibinfo {year} {2020})}\BibitemShut
  {NoStop}%
\bibitem [{\citenamefont {Korzh}\ \emph {et~al.}(2015)\citenamefont {Korzh},
  \citenamefont {Lim}, \citenamefont {Houlmann}, \citenamefont {Gisin},
  \citenamefont {Li}, \citenamefont {Nolan}, \citenamefont {Sanguinetti},
  \citenamefont {Thew},\ and\ \citenamefont {Zbinden}}]{Korzh2015}%
  \BibitemOpen
  \bibfield  {author} {\bibinfo {author} {\bibfnamefont {B.}~\bibnamefont
  {Korzh}}, \bibinfo {author} {\bibfnamefont {C.~C.~W.}\ \bibnamefont {Lim}},
  \bibinfo {author} {\bibfnamefont {R.}~\bibnamefont {Houlmann}}, \bibinfo
  {author} {\bibfnamefont {N.}~\bibnamefont {Gisin}}, \bibinfo {author}
  {\bibfnamefont {M.~J.}\ \bibnamefont {Li}}, \bibinfo {author} {\bibfnamefont
  {D.}~\bibnamefont {Nolan}}, \bibinfo {author} {\bibfnamefont
  {B.}~\bibnamefont {Sanguinetti}}, \bibinfo {author} {\bibfnamefont
  {R.}~\bibnamefont {Thew}},\ and\ \bibinfo {author} {\bibfnamefont
  {H.}~\bibnamefont {Zbinden}},\ }\href
  {https://doi.org/10.1038/nphoton.2014.327} {\bibfield  {journal} {\bibinfo
  {journal} {Nat. Photonics}\ }\textbf {\bibinfo {volume} {9}},\ \bibinfo
  {pages} {163} (\bibinfo {year} {2015})}\BibitemShut {NoStop}%
\bibitem [{\citenamefont {Dynes}\ \emph {et~al.}(2016)\citenamefont {Dynes},
  \citenamefont {Tam}, \citenamefont {Plews}, \citenamefont {Fr{\"{o}}hlich},
  \citenamefont {Sharpe}, \citenamefont {Lucamarini}, \citenamefont {Yuan},
  \citenamefont {Radig}, \citenamefont {Straw}, \citenamefont {Edwards},\ and\
  \citenamefont {Shields}}]{Dynes2016}%
  \BibitemOpen
  \bibfield  {author} {\bibinfo {author} {\bibfnamefont {J.~F.}\ \bibnamefont
  {Dynes}}, \bibinfo {author} {\bibfnamefont {W.~W.}\ \bibnamefont {Tam}},
  \bibinfo {author} {\bibfnamefont {A.}~\bibnamefont {Plews}}, \bibinfo
  {author} {\bibfnamefont {B.}~\bibnamefont {Fr{\"{o}}hlich}}, \bibinfo
  {author} {\bibfnamefont {A.~W.}\ \bibnamefont {Sharpe}}, \bibinfo {author}
  {\bibfnamefont {M.}~\bibnamefont {Lucamarini}}, \bibinfo {author}
  {\bibfnamefont {Z.}~\bibnamefont {Yuan}}, \bibinfo {author} {\bibfnamefont
  {C.}~\bibnamefont {Radig}}, \bibinfo {author} {\bibfnamefont
  {A.}~\bibnamefont {Straw}}, \bibinfo {author} {\bibfnamefont
  {T.}~\bibnamefont {Edwards}},\ and\ \bibinfo {author} {\bibfnamefont {A.~J.}\
  \bibnamefont {Shields}},\ }\href {https://doi.org/10.1038/srep35149}
  {\bibfield  {journal} {\bibinfo  {journal} {Sci. Rep.}\ }\textbf {\bibinfo
  {volume} {6}},\ \bibinfo {pages} {1} (\bibinfo {year} {2016})}\BibitemShut
  {NoStop}%
\bibitem [{\citenamefont {Liu}\ and\ \citenamefont {Yin}(2019)}]{Liu19}%
  \BibitemOpen
  \bibfield  {author} {\bibinfo {author} {\bibfnamefont {P.}~\bibnamefont
  {Liu}}\ and\ \bibinfo {author} {\bibfnamefont {H.-L.}\ \bibnamefont {Yin}},\
  }\href {https://doi.org/10.1364/OSAC.2.002883} {\bibfield  {journal}
  {\bibinfo  {journal} {OSA Continuum}\ }\textbf {\bibinfo {volume} {2}},\
  \bibinfo {pages} {2883} (\bibinfo {year} {2019})}\BibitemShut {NoStop}%
\bibitem [{\citenamefont {Stanco}\ \emph {et~al.}(2022)\citenamefont {Stanco},
  \citenamefont {Santagiustina}, \citenamefont {Calderaro}, \citenamefont
  {Avesani}, \citenamefont {Bertapelle}, \citenamefont {Dequal}, \citenamefont
  {Vallone},\ and\ \citenamefont {Villoresi}}]{Stanco2022}%
  \BibitemOpen
  \bibfield  {author} {\bibinfo {author} {\bibfnamefont {A.}~\bibnamefont
  {Stanco}}, \bibinfo {author} {\bibfnamefont {F.~B.~L.}\ \bibnamefont
  {Santagiustina}}, \bibinfo {author} {\bibfnamefont {L.}~\bibnamefont
  {Calderaro}}, \bibinfo {author} {\bibfnamefont {M.}~\bibnamefont {Avesani}},
  \bibinfo {author} {\bibfnamefont {T.}~\bibnamefont {Bertapelle}}, \bibinfo
  {author} {\bibfnamefont {D.}~\bibnamefont {Dequal}}, \bibinfo {author}
  {\bibfnamefont {G.}~\bibnamefont {Vallone}},\ and\ \bibinfo {author}
  {\bibfnamefont {P.}~\bibnamefont {Villoresi}},\ }\href
  {https://doi.org/10.1109/TQE.2022.3143997} {\bibfield  {journal} {\bibinfo
  {journal} {IEEE Transactions on Quantum Engineering}\ }\textbf {\bibinfo
  {volume} {3}},\ \bibinfo {pages} {1} (\bibinfo {year} {2022})}\BibitemShut
  {NoStop}%
\bibitem [{\citenamefont {Roberts}\ \emph {et~al.}(2018)\citenamefont
  {Roberts}, \citenamefont {Pittaluga}, \citenamefont {Minder}, \citenamefont
  {Lucamarini}, \citenamefont {Dynes}, \citenamefont {Yuan},\ and\
  \citenamefont {Shields}}]{Roberts2018}%
  \BibitemOpen
  \bibfield  {author} {\bibinfo {author} {\bibfnamefont {G.~L.}\ \bibnamefont
  {Roberts}}, \bibinfo {author} {\bibfnamefont {M.}~\bibnamefont {Pittaluga}},
  \bibinfo {author} {\bibfnamefont {M.}~\bibnamefont {Minder}}, \bibinfo
  {author} {\bibfnamefont {M.}~\bibnamefont {Lucamarini}}, \bibinfo {author}
  {\bibfnamefont {J.~F.}\ \bibnamefont {Dynes}}, \bibinfo {author}
  {\bibfnamefont {Z.~L.}\ \bibnamefont {Yuan}},\ and\ \bibinfo {author}
  {\bibfnamefont {A.~J.}\ \bibnamefont {Shields}},\ }\href
  {https://doi.org/10.1364/OL.43.005110} {\bibfield  {journal} {\bibinfo
  {journal} {Opt. Lett.}\ }\textbf {\bibinfo {volume} {43}},\ \bibinfo {pages}
  {5110} (\bibinfo {year} {2018})}\BibitemShut {NoStop}%
\bibitem [{\citenamefont {Avesani}\ \emph {et~al.}(2020)\citenamefont
  {Avesani}, \citenamefont {Agnesi}, \citenamefont {Stanco}, \citenamefont
  {Vallone},\ and\ \citenamefont {Villoresi}}]{Avesani2020}%
  \BibitemOpen
  \bibfield  {author} {\bibinfo {author} {\bibfnamefont {M.}~\bibnamefont
  {Avesani}}, \bibinfo {author} {\bibfnamefont {C.}~\bibnamefont {Agnesi}},
  \bibinfo {author} {\bibfnamefont {A.}~\bibnamefont {Stanco}}, \bibinfo
  {author} {\bibfnamefont {G.}~\bibnamefont {Vallone}},\ and\ \bibinfo {author}
  {\bibfnamefont {P.}~\bibnamefont {Villoresi}},\ }\href
  {https://doi.org/10.1364/OL.396412} {\bibfield  {journal} {\bibinfo
  {journal} {Opt. Lett.}\ }\textbf {\bibinfo {volume} {45}},\ \bibinfo {pages}
  {4706} (\bibinfo {year} {2020})}\BibitemShut {NoStop}%
\bibitem [{\citenamefont {Gr{\"{u}}nenfelder}\ \emph
  {et~al.}(2018)\citenamefont {Gr{\"{u}}nenfelder}, \citenamefont {Boaron},
  \citenamefont {Rusca}, \citenamefont {Martin},\ and\ \citenamefont
  {Zbinden}}]{Grunenfelder2018}%
  \BibitemOpen
  \bibfield  {author} {\bibinfo {author} {\bibfnamefont {F.}~\bibnamefont
  {Gr{\"{u}}nenfelder}}, \bibinfo {author} {\bibfnamefont {A.}~\bibnamefont
  {Boaron}}, \bibinfo {author} {\bibfnamefont {D.}~\bibnamefont {Rusca}},
  \bibinfo {author} {\bibfnamefont {A.}~\bibnamefont {Martin}},\ and\ \bibinfo
  {author} {\bibfnamefont {H.}~\bibnamefont {Zbinden}},\ }\href
  {https://doi.org/10.1063/1.5016931} {\bibfield  {journal} {\bibinfo
  {journal} {Appl. Phys. Lett.}\ }\textbf {\bibinfo {volume} {112}},\ \bibinfo
  {pages} {051108} (\bibinfo {year} {2018})}\BibitemShut {NoStop}%
\bibitem [{\citenamefont {Agnesi}\ \emph {et~al.}(2020)\citenamefont {Agnesi},
  \citenamefont {Avesani}, \citenamefont {Calderaro}, \citenamefont {Stanco},
  \citenamefont {Foletto}, \citenamefont {Zahidy}, \citenamefont {Scriminich},
  \citenamefont {Vedovato}, \citenamefont {Vallone},\ and\ \citenamefont
  {Villoresi}}]{Agnesi2020}%
  \BibitemOpen
  \bibfield  {author} {\bibinfo {author} {\bibfnamefont {C.}~\bibnamefont
  {Agnesi}}, \bibinfo {author} {\bibfnamefont {M.}~\bibnamefont {Avesani}},
  \bibinfo {author} {\bibfnamefont {L.}~\bibnamefont {Calderaro}}, \bibinfo
  {author} {\bibfnamefont {A.}~\bibnamefont {Stanco}}, \bibinfo {author}
  {\bibfnamefont {G.}~\bibnamefont {Foletto}}, \bibinfo {author} {\bibfnamefont
  {M.}~\bibnamefont {Zahidy}}, \bibinfo {author} {\bibfnamefont
  {A.}~\bibnamefont {Scriminich}}, \bibinfo {author} {\bibfnamefont
  {F.}~\bibnamefont {Vedovato}}, \bibinfo {author} {\bibfnamefont
  {G.}~\bibnamefont {Vallone}},\ and\ \bibinfo {author} {\bibfnamefont
  {P.}~\bibnamefont {Villoresi}},\ }\href
  {https://doi.org/10.1364/OPTICA.381013} {\bibfield  {journal} {\bibinfo
  {journal} {Optica}\ }\textbf {\bibinfo {volume} {7}},\ \bibinfo {pages} {284}
  (\bibinfo {year} {2020})}\BibitemShut {NoStop}%
\bibitem [{\citenamefont {Xilinx}()}]{xilGTH}%
  \BibitemOpen
  \bibfield  {author} {\bibinfo {author} {\bibnamefont {Xilinx}},\ }\href
  {http://xilinx.eetrend.com/files/2020-01/wen_zhang_/100046860-87900-ug576-ultrascale-gth-transceivers.pdf}
  {\emph {\bibinfo {title} {UltraScale Architecture GTH Transceivers User
  Guide, UG576}}}\BibitemShut {NoStop}%
\bibitem [{\citenamefont {Scriminich}\ \emph {et~al.}(2022)\citenamefont
  {Scriminich}, \citenamefont {Foletto}, \citenamefont {Picciariello},
  \citenamefont {Stanco}, \citenamefont {Vallone}, \citenamefont {Villoresi},\
  and\ \citenamefont {Vedovato}}]{Scriminich2022}%
  \BibitemOpen
  \bibfield  {author} {\bibinfo {author} {\bibfnamefont {A.}~\bibnamefont
  {Scriminich}}, \bibinfo {author} {\bibfnamefont {G.}~\bibnamefont {Foletto}},
  \bibinfo {author} {\bibfnamefont {F.}~\bibnamefont {Picciariello}}, \bibinfo
  {author} {\bibfnamefont {A.}~\bibnamefont {Stanco}}, \bibinfo {author}
  {\bibfnamefont {G.}~\bibnamefont {Vallone}}, \bibinfo {author} {\bibfnamefont
  {P.}~\bibnamefont {Villoresi}},\ and\ \bibinfo {author} {\bibfnamefont
  {F.}~\bibnamefont {Vedovato}},\ }\href
  {https://doi.org/10.1088/2058-9565/ac8760} {\bibfield  {journal} {\bibinfo
  {journal} {Quantum Sci. Technol.}\ }\textbf {\bibinfo {volume} {7}},\
  \bibinfo {pages} {045029} (\bibinfo {year} {2022})}\BibitemShut {NoStop}%
\end{thebibliography}%

\end{document}